\documentclass[conference]{IEEEtran}
\IEEEoverridecommandlockouts
\usepackage{graphicx}
\usepackage{epstopdf}
\usepackage{cite}
\usepackage{amsmath,amssymb,amsfonts}
\usepackage{algorithmic}
\usepackage{textcomp}
\usepackage{xcolor}
\usepackage{bigints}
\usepackage{mathtools}
\usepackage{gensymb}
\usepackage{multirow}
\usepackage{siunitx}

\DeclareSIUnit{\belmilliwatt}{Bm}
\DeclareSIUnit{\dBm}{\deci\belmilliwatt}

\begin{document}

\title{Line-of-Sight Probability for Outdoor-to-Indoor UAV-Assisted Emergency Networks\\

}



\author{~Gaurav~Duggal, 
~R.~Michael~Buehrer, ~Nishith~Tripathi and ~Jeffrey~H.~Reed


\thanks{G. Duggal, R. M.  Buehrer, N. Tripathi and J.H. Reed  are with Wireless@VT,  Bradley Department of Electrical and Computer Engineering, Virginia Tech,  Blacksburg,
VA, 24061, USA. Email: \{gduggal, rbuehrer, nishith, reedjh\}@vt.edu. The support of AFRL through grant FA8750-20-2-0504, NIST PSCR PIAP through grant: 70NANB22H070 and NSF through grant: CNS-1923807" is gratefully acknowledged.}}
\maketitle



\begin{abstract}
For emergency response scenarios like firefighting in urban environments, there is a need to both localize emergency responders inside the building and also support a high bandwidth communication link between the responders and a command-and-control center. The emergency networks for such scenarios can be established with the quick deployment of Unmanned Aerial Vehicles (UAVs). Further, the 3D mobility of UAVs can be leveraged to improve the quality of the wireless link by maneuvering them into advantageous locations. This has motivated recent propagation measurement campaigns to study low-altitude air-to-ground channels in both 5G-sub6 GHz and 5G-mmWave bands. In this paper, we develop a model for the link in a UAV-assisted emergency location and/or communication system. Specifically, given the importance of Line-of-Sight (LoS) links in localization as well as mmWave communication, we derive a closed-form expression for the LoS probability.  This probability  is parameterized by the UAV base station location, the size of the building, and the size of the window that offers the best propagation path. An expression for coverage probability is also derived. 
 The LoS probability and coverage probabilities derived in this paper can be used to analyze the outdoor UAV-to-indoor propagation environment to determine optimal UAV positioning and the number of UAVs needed to achieve the desired performance of the emergency network.
\end{abstract}


\section{Introduction}
\par
For emergency response scenarios like firefighting and rescue, the incident scene is often served by a variety of emergency personnel such as police, firefighters, and medical personnel. In these events, there is a need to localize emergency responders inside the building and often to transmit a real-time high-quality video feed to a central location outside the building for  situational awareness \cite{merwadayuav},\cite{merwaday2015uav}. There is a benefit in offering bi-directional high bandwidth wireless connectivity since it can be used to provide responders with real-time 3-D information. Further, the vital signs or other critical health information concerning injured survivors could be transmitted to medical staff prior to extraction. 
\par
In these emergency scenarios,  the wireless signal propagation environment is between a Base Station (BS) located outside the building to a Mobile Station (MS) located inside. Both the fading and path loss characteristics of the signal propagation are dependent on the material properties and the position of the BS and MS. 
\par
Prior work has explored the use of UAVs to aid in the deployment of emergency wireless networks to reduce reliance on existing cellular infrastructure \cite{merwadayuav},\cite{merwaday2015uav}. One reason for this is that the 3D mobility of the UAVs can be leveraged to maneuver UAV-mounted BSs into advantageous positions that offer higher-quality wireless links. Recently, outdoor-to-indoor propagation measurement campaigns have been conducted using UAVs in both 5G-sub6 GHz bands ($1\si{\GHz}-6\si{\GHz}$) \cite{saito20194} and 5G mmWave bands ($24\si{\GHz}-43.5\si{\GHz}$) \cite{fuschini2022multi}. In the latter studies, researchers have noted that the path loss is primarily dependent on the building materials and the surface area of the windows. This motivates an analytical study to investigate the effect of windows, building geometry, and the relative position of the BS  on the link quality, which is a key objective of the paper.  

One of the conditions that affects the link quality is whether the direct path between the transmitter and receiver is blocked or not \cite{khawaja2019survey}. For system-level analyses, the metric that quantifies this is the LoS probability. The importance of LoS probability to system performance is seen in its inclusion in 3GPP models \cite{3gpp_channel_model}.  In general there are three main approaches that have been adopted to characterize  LoS probability: i) empirical methods that use curve fitting from real-world measurements; ii) map-based methods that employ tools like ray-tracing to model electromagnetic effects and iii) stochastic methods that rely on an appropriate  model of the environment under investigation. While each of these methods have relative advantages, the third approach is particularly useful in obtaining tractable models that can be incorporated into larger system-level analyses. On the other hand, the first two are largely confined to specific environments and are, therefore, harder to generalize. 
\par
The concept of Fresnel zones has been used extensively to derive the LoS probability in both outdoor and indoor environments, but to the best of our knowledge, not indoor-to-outdoor environments. Feng {\em et al} \cite{feng2006wlcp2} considers the diffraction losses around buildings and uses the intrusion percentage  of an obstacle inside the first Fresnel zone as an LoS condition. Liu {\em et al} \cite{liu2018analysis} uses the same LoS condition in various outdoor scenarios to derive LoS probability and shows that the approach  agrees with the empirical models in the 3GPP standards. Hmamouche {\em et al} \cite{hmamouche2022fresnel} use the two-ray propagation model and the Knife Edge Diffraction (KED) model to analytically and numerically demonstrate the validity of defining the LoS condition based on  the intrusion ratio of an obstruction in the first Fresnel zone. Each of these works derives the LoS probability and shows its effects on the quality of the wireless link. The same LoS condition has been used for new scenarios like in Jarvelinen {\em et al} \cite{jarvelainen2016evaluation},\cite{jarvelainen201470} where they have used point cloud data in their model to obtain the LoS probability for i) an urban open square, ii) an indoor shopping mall and iii) indoor office scenarios. Most recently Ulloa {\em et al} \cite{ulloa2020millimeter} have modeled the LoS probability in metro carriages.
\par
In this paper, we study the LoS probability for a specific outdoor-to-indoor setting in which a UAV located outside a building is attempting to establish an LoS link with a user located inside the building through a window of specific dimensions. Even though this setup may appear simple at first glance, it must be noted that it is a canonical construction that can be easily extended to include multiple windows on a given floor of a building, as well as to multi-floor scenarios. Using the idea of the Fresnel zone, we first establish the condition for the existence of LoS in this setting, using which we derive the LoS probability by assuming that the user is uniformly distributed in the room. In order to make this model useful for further system-level analyses, our main goal is to derive closed-form expressions even if we need to make careful approximations along the way.

\begin{table}[!t]
\caption{Model parameters with typical values}
\centering
\begin{tabular}{|c|c|c|}
\hline
\textbf{Parameter} & \textbf{Notation} & \textbf{Range}\\
\hline
Room dim. & $L_r$ & 10m-40m\\
\hline
Window dim. &$L_w$ & 1m-5m \\
\hline
BS dist. &$d_a$ & 2m-100m \\
\hline
BS angle & $\theta$ & $-90\degree$ - $90\degree$  \\
\hline
\end{tabular}
\label{table:model_params}
\end{table}

\section{System Model and Problem formulation}
In this section, we first introduce the system model and then using the Knife Edge Diffraction (KED) model, show the validity of using the percentage intrusion of the obstacle in the first Fresnel zone as a LoS condition for our outdoor-to-indoor link. The Los transmission is shown to depend on the relative position of the BS outside the building. We also derive the LoS probability as a function of the model parameters.

\subsection{System Model}
\begin{figure}[t]
\centering
         \includegraphics[width=0.5\linewidth]{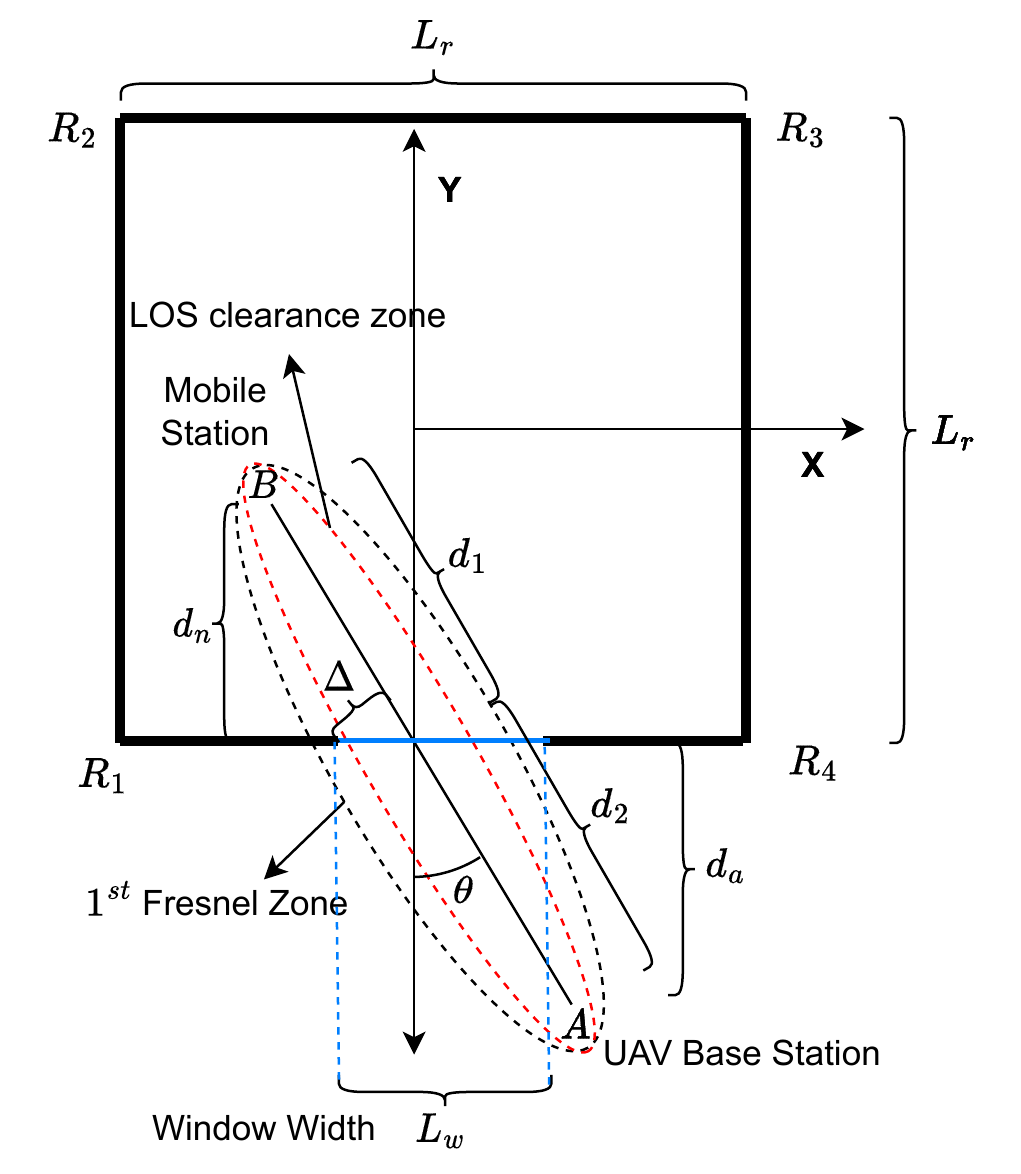}
         \caption{Top view for the outdoor-to-indoor emergency network scenario with the UAV-mounted BS at point $A$, the MS at point $B$ inside a building represented by $\square R_1R_2R_3R_4$.}
         \label{fig:room_model}
\end{figure}

We assume that based on the inputs from the emergency responders inside the building, the UAV-mounted BS can be maneuvered to a given floor. We then consider the cross-section of a large room as in Fig.\ref{fig:room_model}, with the BS located at point $A$ outside the room and the MS located somewhere inside the room. The square $\square R_1R_2R_3R_4$ represents the room of floor dimensions $L_r \times L_r $ with the exterior walls made out of concrete. There is a standard non-metallic glass window of width $L_w$ located at the center of one of the walls of the room. Due to the high attenuation offered by concrete, it is assumed to be a blocker. On the other hand, the non-metallic glass windows offering lower attenuation are considered to be transparent to radio signal propagation. The BS location is defined by distance $d_a$ from the room, at an angle $\theta$ from the normal to the window located at the center of the window. For a fixed BS location, window, and room geometry, the MS could be uniformly located anywhere inside the room. Some percentage of MS locations will be LoS depending on if they satisfy the LoS condition whereas the other locations will be NLoS.

\subsection{Knife Edge Diffraction (KED)}
For this scenario, we employ the KED model to numerically validate the LoS condition that the ratio of the intrusion of the obstacle into the first Fresnel zone governs the transitions from LoS to NLoS. When blockages approach the optical line joining a transmitter and receiver, or even cross it, the KED model can approximate the diffraction losses at the receiver due to diffraction around the obstacle \cite{rappaport1996wireless}. In Fig. \ref{fig:room_model} the window edge lies at distance $\Delta$ and is an obstacle to signal propagation from the transmitter at point $A$ to the receiver at point $B$. The equation governing the diffraction losses at point $B$ due to an obstruction at distance $\Delta$ from the direct path joining the transmitter and receiver is given by
\begin{multline*}
   PL_{diff} =  \\ -20\log \left({\frac{\sqrt{(1-C(v)-S(v))^2+(C(v)-S(v))^2}}{2}}\right).   
\end{multline*}
Here, $C(v)$ and $S(v)$ are the real and imaginary parts of the complex Fresnel integral $F_c(v)=\int_{0}^{v}  e^{(j\frac{\pi s^2}{2})} \,{\rm d}s$. Further, $v$ is a dimensionless parameter and is given by \eqref{eq:dimensionless_fresnel_integral_paraneter} expressed as a function of the distances $d_1$, $d_2$ as in Fig. \ref{fig:room_model}.
\begin{equation}
    v = \Delta \sqrt{\frac{2}{\lambda}\left(\frac{1}{d_1}+\frac{1}{d_2}\right)}
    \label{eq:dimensionless_fresnel_integral_paraneter}
\end{equation}
Note that $\Delta$ is measured from the straight line joining the transmitter and receiver as shown in Fig. \ref{fig:room_model}. It is positive if the obstruction does not obstruct the direct path and is negative if it does. In our case, for a fixed BS location and for various MS locations inside the room, the obstruction distance $\Delta$ changes and this affects the path loss experienced at the MS. 
\par
\begin{figure}[!t]
\centering
         \includegraphics[width=0.6\linewidth]{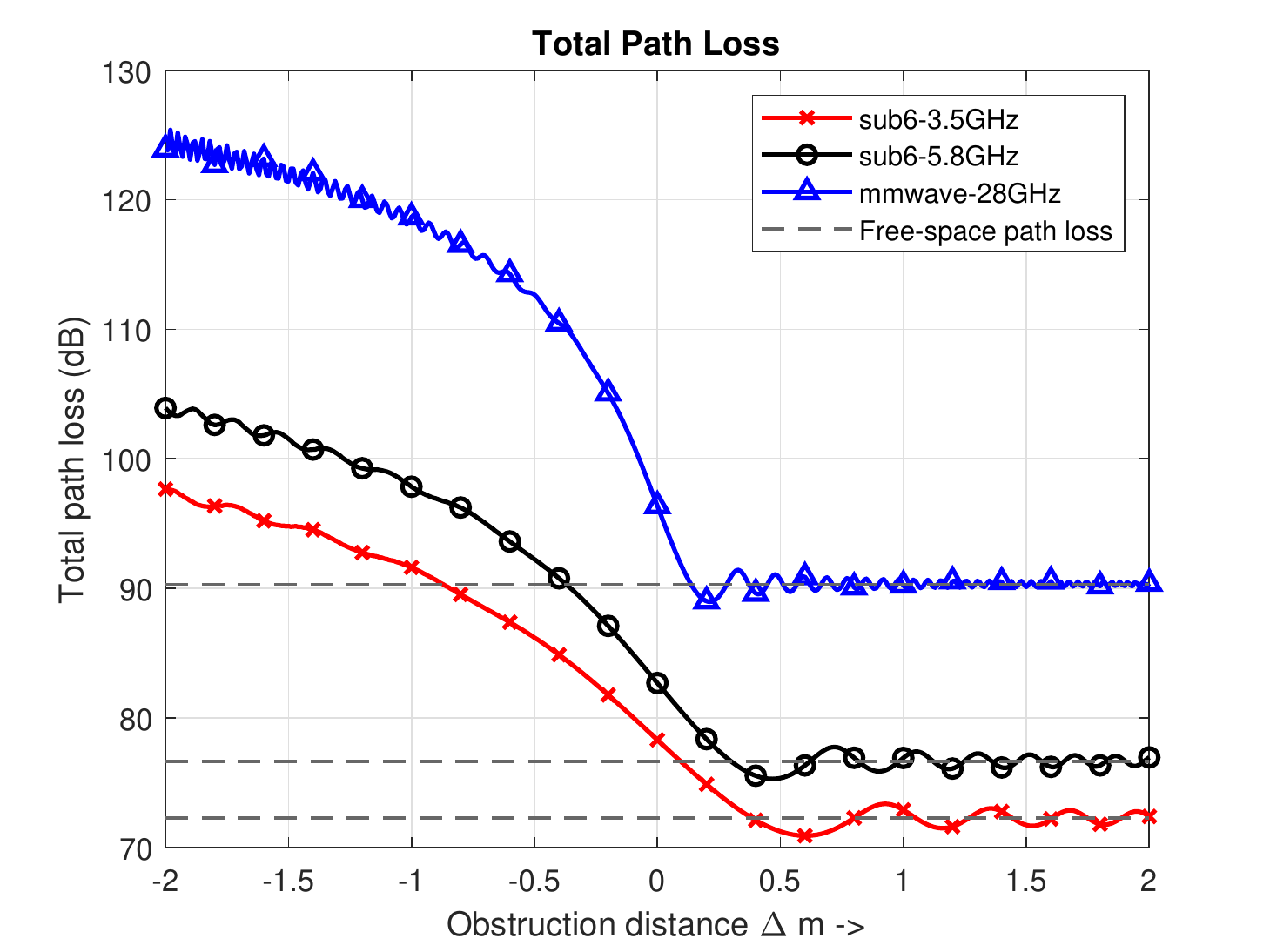}
         \caption{Path loss as a function of the intrusion distance of an obstacle into the first Fresnel zone region for 5G-sub6 and 5G-mmWave frequencies for $d_1 = 8m$, $d_2=20m$.}
         \label{fig:Path_loss}
\end{figure}

As an example, for the geometry in Fig. \ref{fig:room_model}, fix distances $d_1$ and $d_2$. Now, for different MS and BS locations constrained by $d_1$ and $d_2$, the window edges form obstacles to the signal propagation between the BS and MS by intruding into the first Fresnel zone. The first Fresnel radius at the point where the windows act as obstructions is given by $r_d=\lambda \sqrt{\frac{d_1d_2}{d_1+d_2}}$. in Fig. \ref{fig:Path_loss}, on setting $d_1=8m$ and $d_2=20m$, we plot the total path loss which is the sum of the free space path loss and diffraction path loss. This path loss is a function of the intrusion distance $\Delta$ for 5G-sub6 and 5G-mmWave bands. It is clear from \ref{fig:Path_loss}, the transition region from LoS to NLoS should be based on the intrusion ratio of the obstacle into the first Fresnel zone region and numerically we set the ratio $\Delta / r_d \approx 0.6$ which is $60\%$ blockage of the first Fresnel zone region. This LoS condition can be visualized in Fig. \ref{fig:room_model} as the red ellipsoidal region lying inside the first Fresnel zone and is called the LoS clearance region. This region needs to be kept clear of obstructions from the window edges for LoS transmission. Hence for LoS transmission, $\Delta \geq 0.6 r_d$ and will be used for further analysis.

\section{LoS probability}
In this section, we define the LoS probability using the LoS condition based on the intrusion ratio of the obstacle into the first Fresnel zone. With the motivation to gather insights on the variation of the LoS probability on the BS location, building size as well as window dimensions, we approximate the LoS probability as a closed-form expression in terms of our model parameters.
\begin{figure}[!ht]
\centering
         \includegraphics[width=0.80\linewidth]{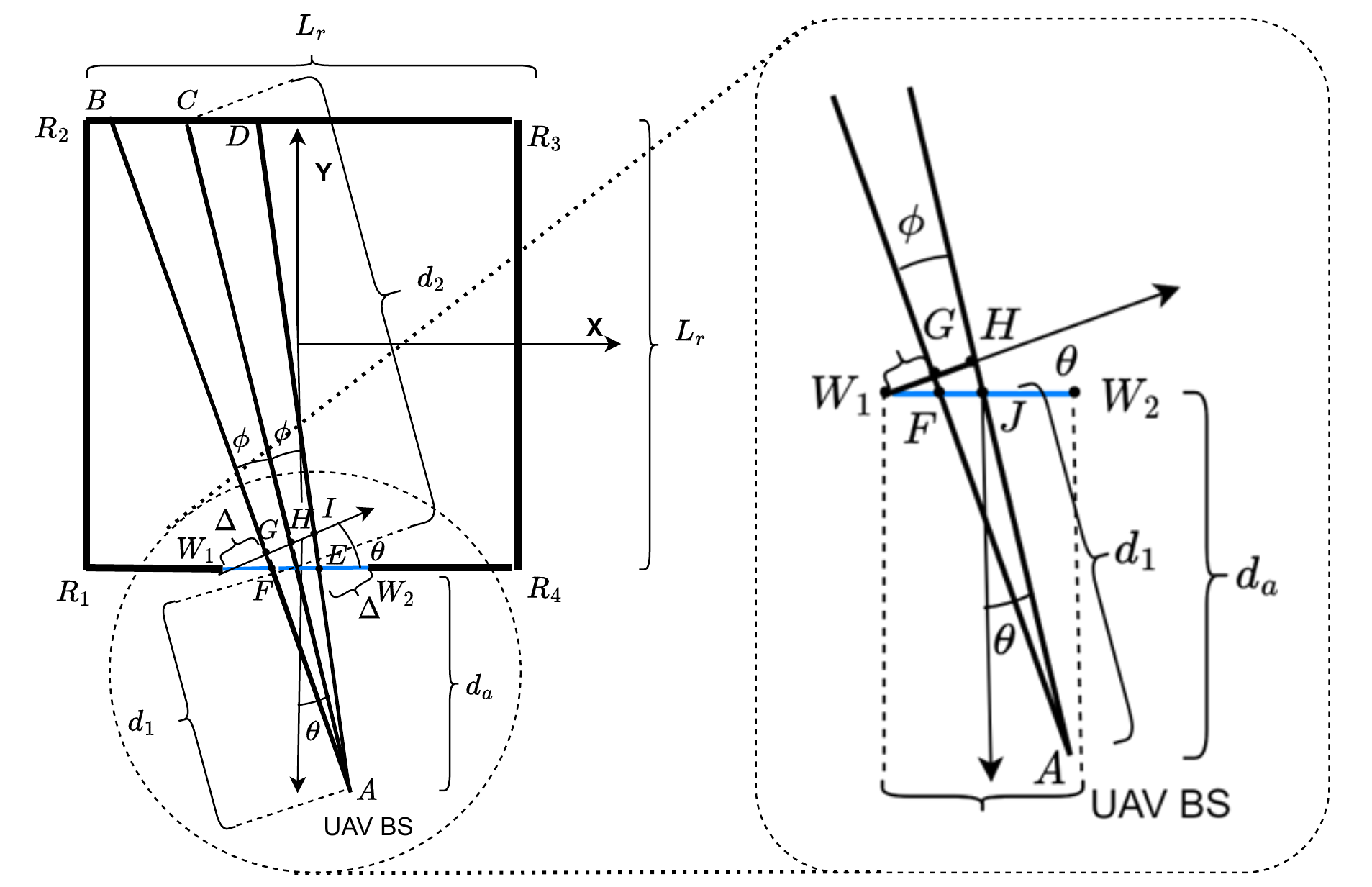}
         \caption{LoS region $\square FBDE$  when the back wall is illuminated by the UAV-BS at point $A$.}
         \label{fig:los_region_back_wall}
\end{figure}
\begin{figure}[!ht]
\centering
         \includegraphics[width=0.6\linewidth]{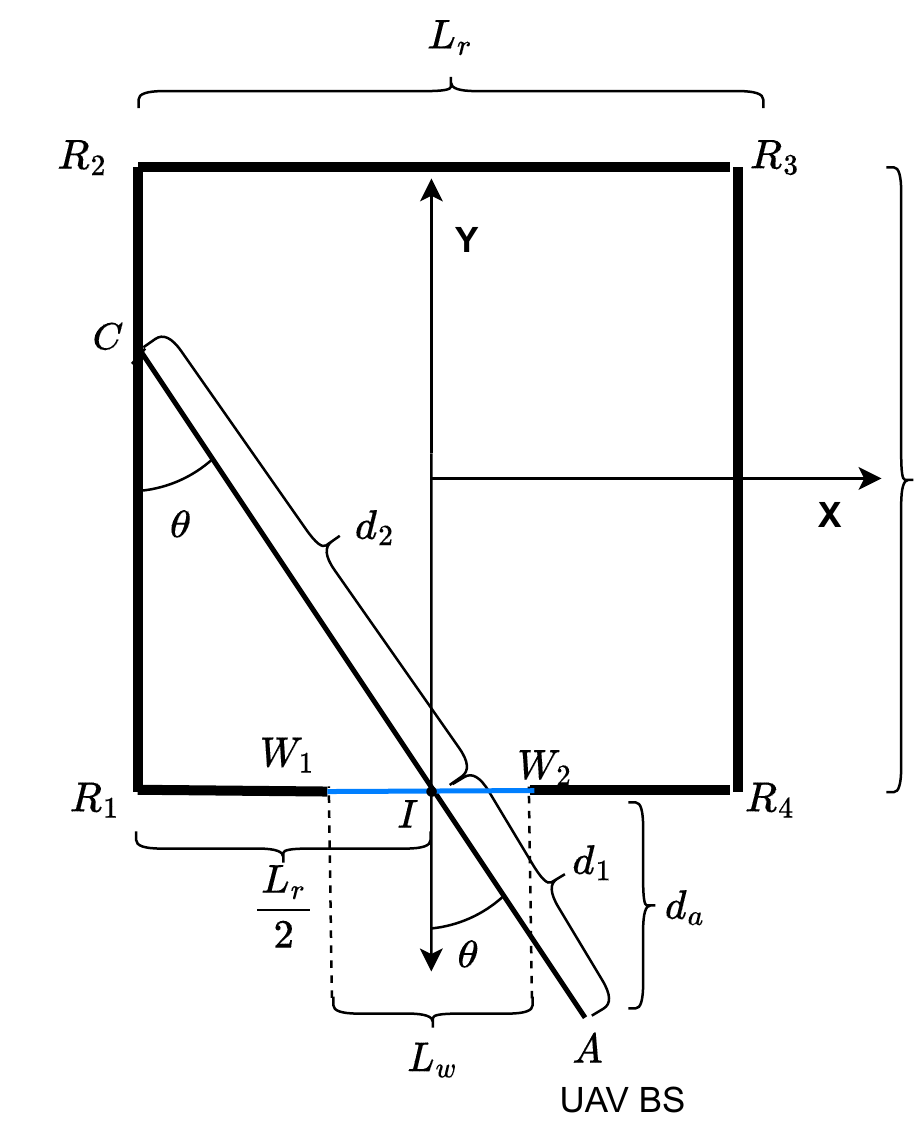}
         \caption{Geometry for the scenario when the side wall is illuminated by the UAV-BS at point $A$.}
         \label{fig:los_region_side_wall}
\end{figure}
\par
The BS is fixed at a point $F$ parameterised by distance $d_a$ and angle $\theta$ as shown in 
Fig. \ref{fig:los_region_back_wall}. Let $B$ and $D$ be the leftmost and rightmost points lying on the back wall which are LoS from the BS. Both points $B$ and $D$ are offset by angle $\phi$ on opposite sides of the line $AC$. Note that the angle subtended by the line $BD$ at the point $A$ is $2\phi$ and as the window dimension $L_w$ is much smaller than both the BS distance and MS distance, $\phi$ is a very small angle hence $AB \approx AD \approx AC = d_2$, $AF \approx AE \approx AI = d_1$. Applying the LoS transmission condition $\Delta \geq 0.6 r_d$ for points $B$ and $D$ observe that all points within the region defined by quadrilateral $\square FBDE$ are LoS and this region is called the LoS region. Hence the LoS probability parameterised by the BS location ($d_a,\theta$) room dimension $L_r$ and window dimension $L_w$ is given by:
\begin{equation}
    P_{los}(\theta,d_a,L_r,L_w) = \frac{area(\square FBDE)}{area(room)} = \frac{area(\square FBDE)}{L_r^2}.
    \label{eq:p_los}
\end{equation}
To obtain the simulation result for the LoS probability, we place a uniform $N\times N$ grid of possible MS locations within the room. The LoS probability can be obtained by $P_{los} = \frac{N_{los}}{N^2}$ where $N_{los}$ is the number of grid locations in the LoS region $\square FBDE$ for a fixed BS location. To get a closed-form expression for analyzing the LoS probability, in Fig. \ref{fig:los_region_back_wall}, we approximate the area of $\square FBDE$ by considering the line segments $BD$ and $FE$ as arcs of radii $d_1+d_2$ and $d_1$ respectively. 
\begin{equation}
\begin{aligned}
    area(\square FBDE) & \approx (d_1+d_2)^2 \phi - d_1^2 \phi \\
\end{aligned}
\label{eq:area_los_region}
\end{equation}
Referring to the zoomed part of Fig. \ref{fig:los_region_back_wall}, the arc approximation is used to obtain $\phi$ in terms of the model parameters with the last step obtained by substitution of the LoS condition.
\begin{equation}
    \begin{aligned}
    &\phi= \frac{GH}{AH}\\[1ex]
    &\approx \frac{\frac{L_w\cos\theta}{2} -\Delta}{d_1}\\[1ex]
    &= \frac{L_w\cos^2 \theta -1.2r_d \cos\theta}{2d_a}\\
    \end{aligned}
    \label{eq:phi_los_region}
\end{equation}
There are two cases for $d_2$ depending on the dimensions of the room. The condition for point B to lie on the back wall is $\theta \in (-\theta_0,\theta_0)$ where $\theta_0=26\degree$ since our room is square. It is straightforward to get $d_2$ in terms of the BS distance $d_a$ and $\theta$ in that case. When point C lies on the side wall as shown in Fig. \ref{fig:los_region_side_wall}, we employ the geometry in $\triangle CR_1I$ to get the formulation for $d_2$.

\begin{figure}[!t]
\centering
         \includegraphics[width=0.60\linewidth]{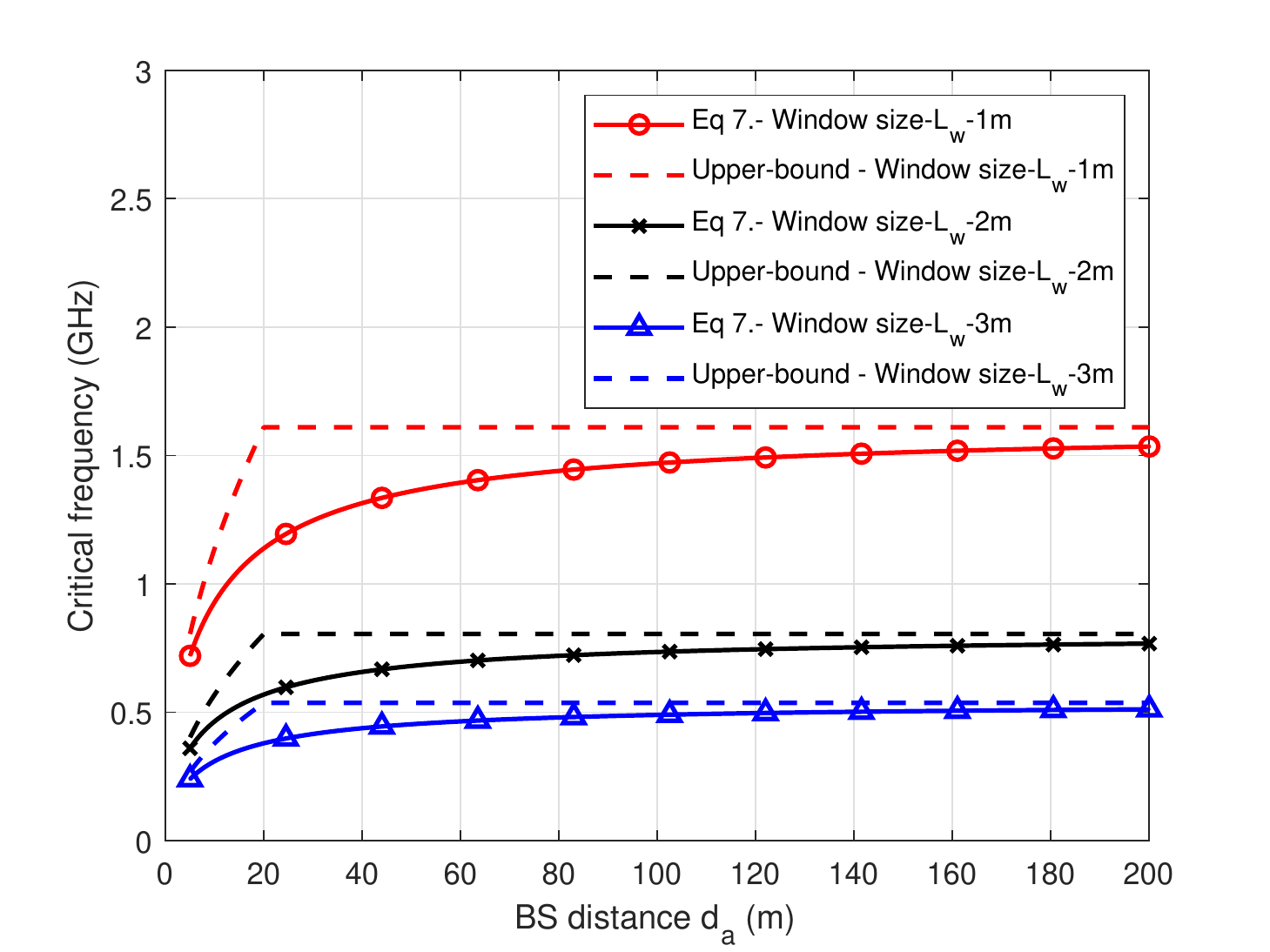}
         \caption{For the frequencies in consideration i.e. in 5G-sub6 bands and 5G-mmWave bands, the regime of operation is above the critical frequency. In the plot the we compare the closed form expression of the critical frequency vs its upper bound for room size $L_r=20m$ and window size $L_w=1m,\;2m,\;3m$. }
         \label{fig:critical_frequency}
\end{figure}

\begin{equation}
\small
d_2=
    \begin{cases}
    \begin{aligned}
        \frac{L_r}{\cos\theta}, & \;\;\;\theta \in (-\theta_0\degree,\theta_0\degree) \\[1ex]
        \left|\frac{L_r}{2\sin\theta}\right|, &   \;\;\;     \theta \in (-90\degree,-\theta_0\degree) \cup (\theta_0\degree,90\degree) \\
    \end{aligned}
    \end{cases}
    \label{eq:d2_los_region}
\end{equation}
Note using $d_1=\frac{d_a}{\cos\theta}$, the closed-form expression for the LoS probability is obtained by substituting \eqref{eq:phi_los_region},\eqref{eq:d2_los_region} in \eqref{eq:area_los_region} and finally in \eqref{eq:p_los}.

\begin{equation}
\small
\begin{aligned}
P_{los}&(\theta,d_a,L_r,L_w) = \\
    &\begin{cases}
         \left(\frac{L_w\cos^2\theta-1.2r_d\cos\theta}{2L_r^2d_a}\right)&\left(\frac{2L_rd_a+L_r^2}{\cos^2\theta}\right),    \\ & \theta \in (-\theta_0\degree,\theta_0\degree) \\[1ex]
        \left(\frac{L_w\cos^2\theta-1.2r_d\cos\theta}{2L_r^2d_a}\right)&\left(\frac{2d_a}{\cos\theta}+\left|\frac{L_r}{2\sin\theta}\right|\right)\left|\frac{L_r}{2\sin\theta}\right| 
        \; \\  &\theta \in (-90\degree,-\theta_0\degree)\cup(\theta_0\degree,90\degree) \\
    \end{cases}
    \end{aligned}
    \label{eq:p_los_closed_form}
\end{equation}
Here $r_d$ is the Fresnel radius at distance $d_1,d_2$.
\section{LoS probability Results}
In this section, we systematically analyze the variation of the LoS probability with the system parameters. Specifically, we make conclusions for both 5G-sub6 bands as well as 5G-mmWave bands. At 5g-mmWave frequencies, we show that the frequency variation becomes negligible. 
\subsection{Critical frequency}
In \eqref{eq:phi_los_region} substituting $\theta=0\degree$, we observe that $\phi$ is dependant on the frequency, and for a particular frequency defined as critical frequency, it becomes zero. On examining \eqref{eq:p_los_closed_form}, for $\theta=0\degree$ and frequency of operation at/or below $f=$ critical frequency, we observe that the LoS probability is zero. We can obtain an expression for the critical frequency by equating \eqref{eq:phi_los_region} to zero to get \eqref{eq:critical_frequency}. The critical frequency is dependent on the BS distance, room dimensions and inversely proportional to the window size. The square root term in the expression for the critical frequency has the BS distance $d_a$ and room dimension $L_r$. This square root term can be upper bounded by the square root of the smaller dimension of the two. From the plot in Fig. \ref{fig:critical_frequency} we conclude for typical model parameters in table \ref{table:model_params}, all of the 5G-sub6 and 5G-mmWave bands lie above the critical frequency, and hence the further analysis will be confined to this regime of operation.  
\begin{equation}
    f = \frac{1.2c}{L_w}\sqrt{\frac{1}{\frac{1}{d_a}+\frac{1}{L_r}}} < \frac{1.2c}{L_w}\sqrt{\min(d_a,L_r)} 
    \label{eq:critical_frequency}
\end{equation}
 
\subsection{LoS probability vs frequency}

\begin{figure}[!t]
\centering
         \includegraphics[width=0.60\linewidth]{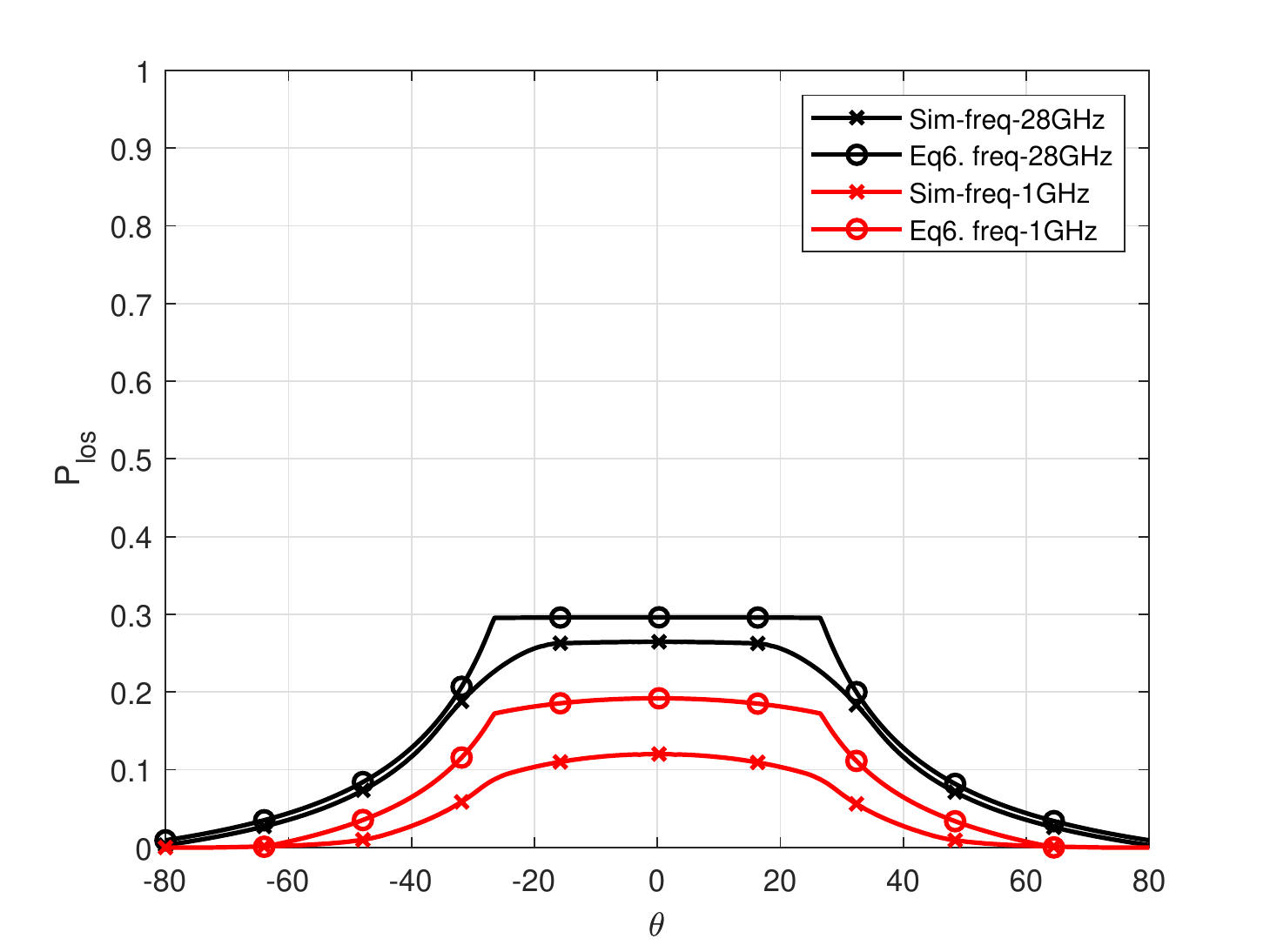}
         \caption{Simulation and closed form expression \eqref{eq:p_los_closed_form} vs BS aspect angle to window $\theta$ for 5G-sub6 band $f=1\si{\GHz}$ and 5G-mmWave band $f=28\si{\GHz}$ for BS distance $d_a=5m$, room dimension $L_r=20m$, window dimensions $L_w=2m$. }
         \label{fig:plos_vs_frequency}
\end{figure}

\begin{figure}[!t]
\centering
         \includegraphics[width=0.60\linewidth]{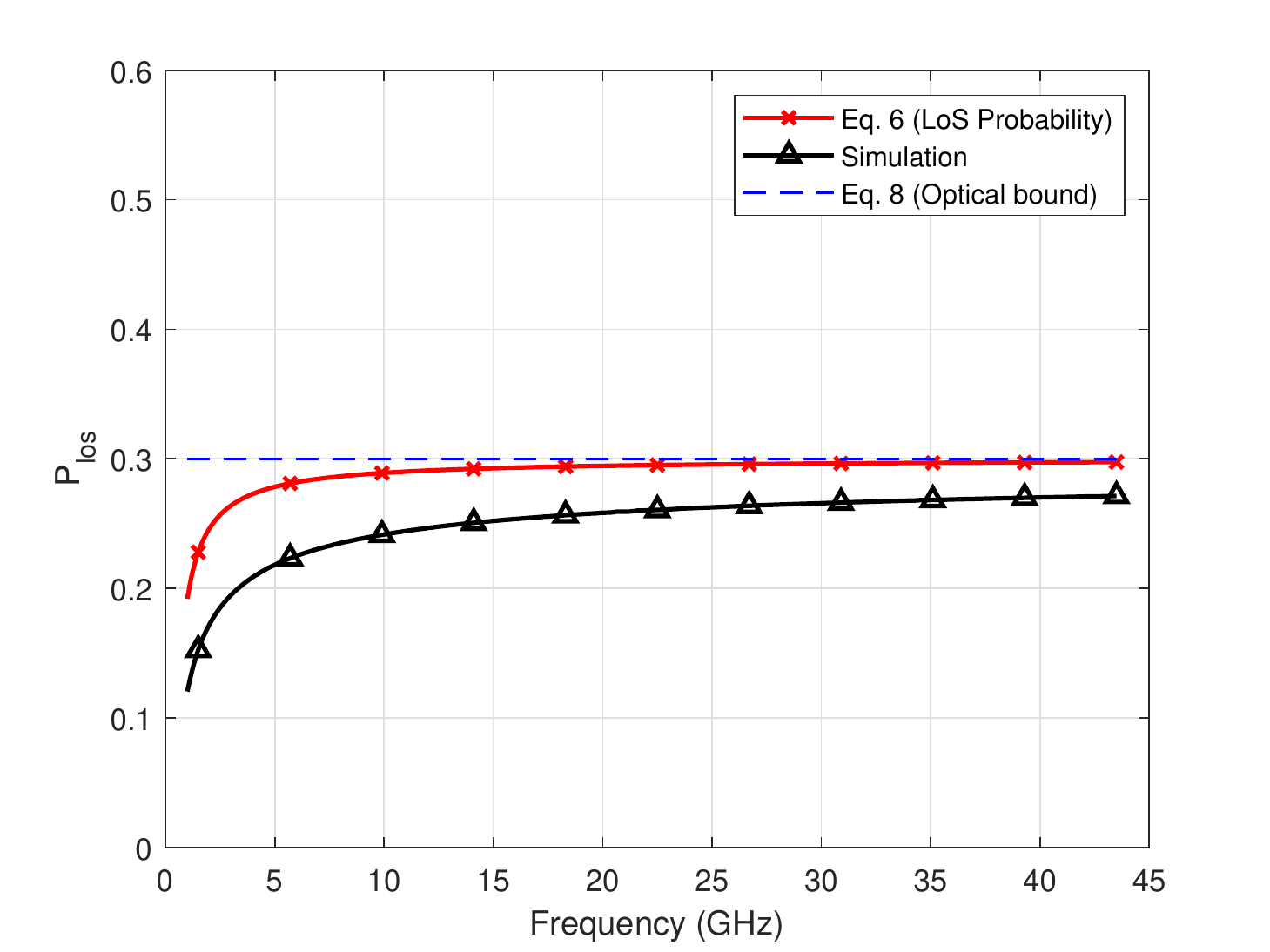}
         \caption{Frequency dependence of the LoS probability \eqref{eq:p_los_closed_form}, Optical bound \eqref{eq:plos_optical_upper_bound},   and simulation for BS distance $d_a=5m$, room dimension $L_r=20m$, window dimensions $L_w=2m$. }
         \label{fig:plos_vs_optical_vs_sim}
\end{figure}

As the frequency of operation is increased, the first Fresnel zone radius reduces leading to an increase in the angle $\phi$. Hence as we move from 5G-sub6 to 5G-mmWave frequencies we expect the LoS probability to increase, as observed in Fig. \ref{fig:plos_vs_frequency}. Hence in order to maximise the LoS probability it is desirable to use 5G-mmWave bands over 5G-sub6 bands. Specifically for 5G-mmWave frequencies ($24\si{\GHz}-43.5\si{\GHz}$), we note that the first Fresnel zone radius can be assumed to be much smaller than the dimension of the window $L_w$. This corresponds to assuming an infinite frequency of operation and in this case, the LoS to NLoS transition happens when the obstructions to the signal propagation just touches the direct path joining the BS and MS. This is equivalent to dropping the Fresnel radius term in \eqref{eq:p_los_closed_form} and by substituting $\theta=0\degree$ we obtain a frequency independent ``optical" approximation for the LoS probability. Observe that in Fig. \ref{fig:plos_vs_optical_vs_sim}, for 5G-mmWave bands, the optical bound and the closed form expression for the LoS probability converge and the simulation is in very close agreement.
\begin{equation}
    P_{los} \approx L_w\left(\frac{1}{L_r}+\frac{1}{2d_a}\right)
    \label{eq:plos_optical_upper_bound}
\end{equation}

\subsection{LoS probability vs Window Size}
From \eqref{eq:p_los_closed_form} as we increase the window size, $\phi$ increases, hence the LoS probability is expected to rise for both 5G-sub6 as well as 5G-mmWave bands. For the latter we can neglect the Fresnel radius leading to a simplified expression for the LoS probability in \eqref{eq:plos_optical_upper_bound}. From this, we conclude that $P_{los}$ is directly proportional to the dimension of the window $L_w$.    
\begin{figure}[!t]
\centering
\includegraphics[width=0.60\linewidth]{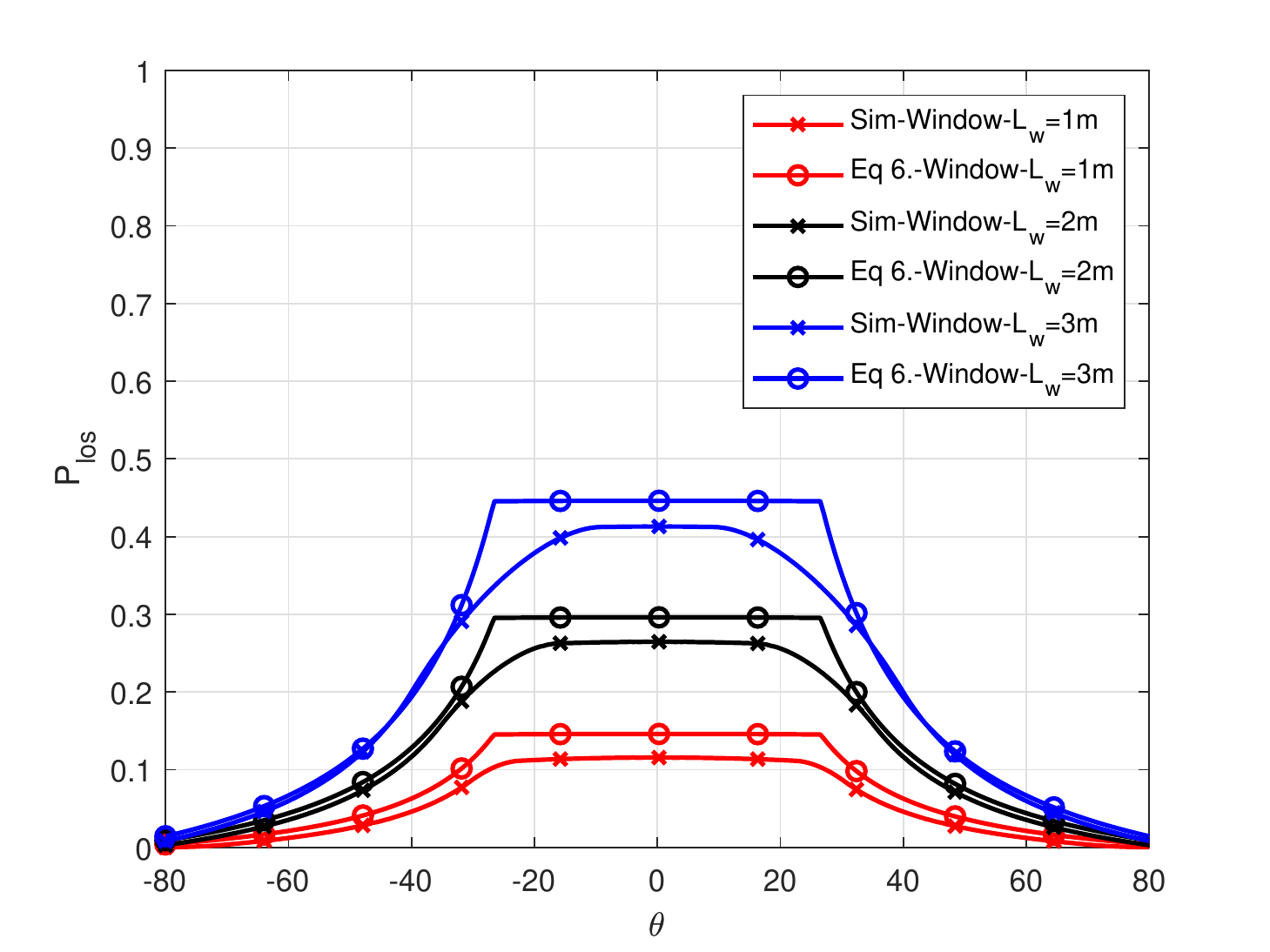}
         \caption{LoS probability is directly proportional to the window size $L_w$ for 5G-mmWave bands. $f=28\si{\GHz}$, BS distance $d_a=5m$, room dimension $L_r=20m$, window dimensions $L_w=1m,2m,3m$. }
         \label{fig:plos_vs_window_size}
\end{figure}
\subsection{LoS probability vs dimension of room and BS distance}
 As the BS moves away from the room i.e. $d_a$ increases, observe that $\phi$ decreases, hence the LoS probability will decrease initially. Now for high frequencies the Fresnel radius can be ignored compared to the window dimension, note that in Fig. \ref{fig:los_region_back_wall} for $\theta=0\degree$, beyond a certain BS station distance, the line segments $FB$ and $ED$ become smaller than $d_a$. Geometrically, for sufficiently large $d_a$, they become parallel to each other. At this point there should be no further decrease in the area of the LoS region and the LoS probability will not decrease further. Equivalently, in \eqref{eq:plos_optical_upper_bound} as $\frac{1}{d_a}$ becomes negligible to $\frac{1}{L_r}$ and it can be ignored. In \eqref{eq:plos_optical_upper_bound}, we can make similar arguments regarding the room dimension $L_r$ and the LoS probability is essentially dominated by the smaller dimension between the BS distance $d_a$ and the room dimension $L_r$.

\begin{figure}[!t]
\centering
\includegraphics[width=0.65\linewidth]
{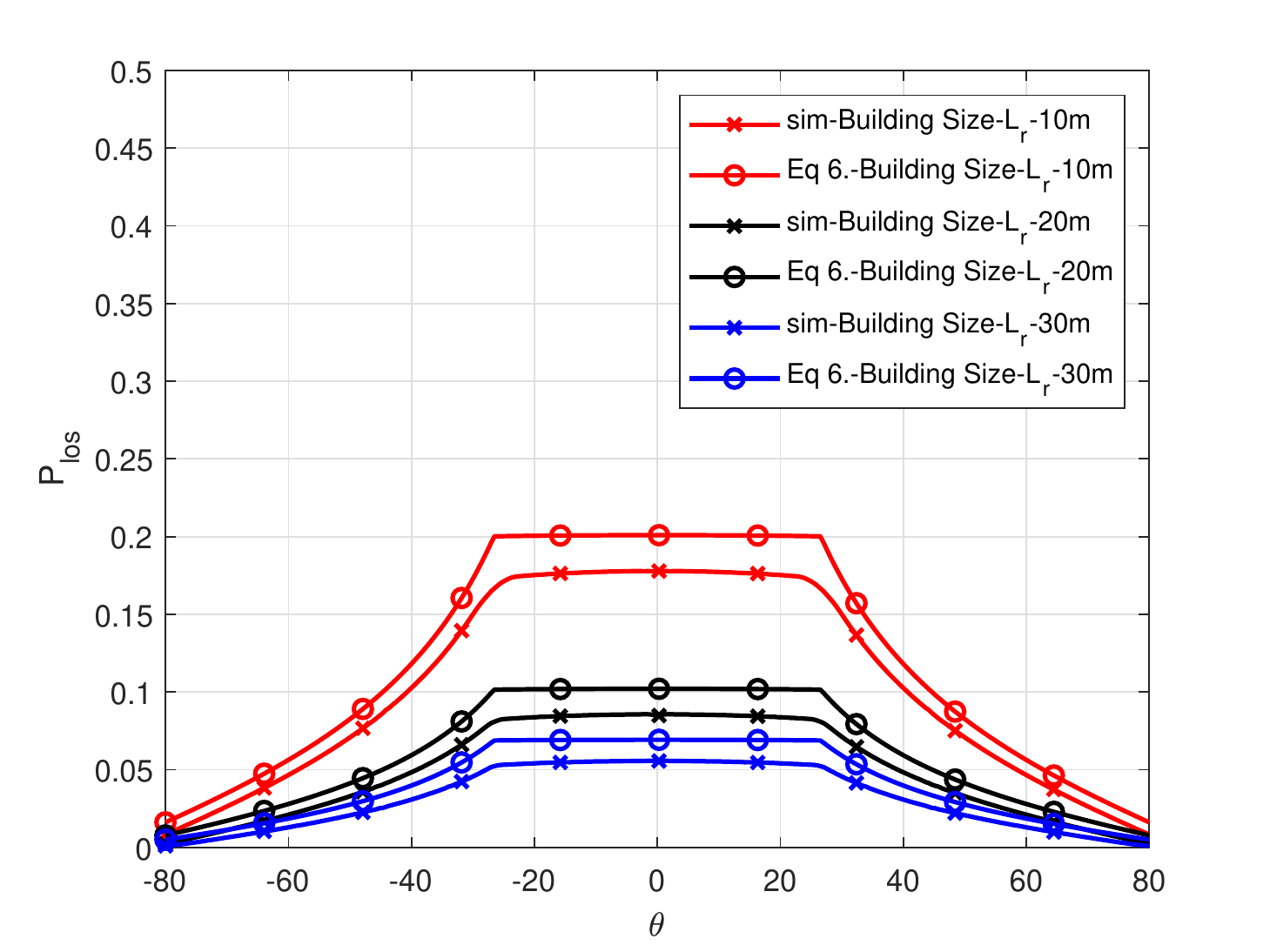}
         \caption{LoS probability is inversely proportional to size of the room for 5G-mmWave bands. $f=28\si{\GHz}$ for BS distance $d_a=5m$, window dimension $L_w=2m$, room dimension $L_r=10m,20m,30m$. }
         \label{fig:plos_vs_room_size}
\end{figure}

\section{Coverage probability}
In this section, we evaluate the quality of the wireless link from one UAV mounted BS to an indoor MS for 28 $\si{\GHz}$ for our scenario in terms of the coverage probability. The coverage probability expression is shown to be a function of the LoS probability. As the wireless signal propagates from the BS to the MS, it experiences both fading as well as path loss. For a fixed BS location the MS location decides if the wireless link is LoS and NLoS. Hence, we need to model both fading and path loss for LoS and NLoS scenarios. To model the fading, we use the Nakagami-m fading model with appropriate use of the parameter $m$. The channel power gain $\gamma$ in case of LoS propagation is Gamma distributed ($~\Gamma(m,\frac{1}{m})$) with $m\approx10$, whereas in case of NLoS propagation it is Rayleigh distributed ($~\Gamma(m,\frac{1}{m})$) with $m\approx 1$. Note that $m=10$ corresponds to Ricean K factor $~13\;\si{\dB}$. Based on the measurement campaigns for $28 \si{GHz}$ in Sun {\em et al} \cite{sun2016millimeter}, the path loss exponent is 1.2 for LoS and 2.9 for NLoS scenarios. The received average signal-to-noise ratio $\gamma$ can be modelled by
\begin{equation}
    \gamma = \frac{\lambda^2}{16\pi^2} \frac{P_t}{d^n}\frac{1}{KTB}.
    \label{eq:received_snr}
\end{equation}
Here, $n$ is the path loss exponent, $\lambda$ is the wavelength and $P_t$ is the transmit power, $KTB$ is the thermal noise floor of the receiver and $d$ is the distance between the transmitter and receiver.  
For a BS at $\theta=0\degree$ all MS locations at distance $d_a+d_n$ from the BS can be approximated to lie on a line segment of dimension $L_r$ equal to the width of the room. Since the distance between the MS and BS is known {\em apriori}, the coverage probability is defined to be the percentage of MS locations lying on the line segment with an SNR above a preset threshold $\gamma_T$.
\par
To get the LoS probability for a fixed BS location and MS location, we observe that the LoS MS locations lying at a distance $d_a+d_n$ from the BS subtend an angle $2\phi$ at the BS. Using the small arc approximation the LoS probability for $\theta=0\degree$ can be written as  
\begin{equation}
    P_{los}= \frac{(d_a+d_n)}{d_n}\frac{(L_w-1.2r_d)}{d_a}.
    \label{eq:Los_probability_distance}
\end{equation}
This can be visualized as the percentage of the LoS MS locations lying on the line segment at a distance $d_a+d_n$ that are in the LoS region. Hence the coverage probability depends on both the LoS probability and the fading parameters and can be expressed as
\begin{equation}
\begin{aligned}
\small
P_{cov}&= \mathbb{P}(\gamma_{los}>\gamma_T, LoS )+\mathbb{P}(\gamma_{nlos}>\gamma_T, NLoS)\\ 
&= \mathbb{P}_{los}(\gamma_{los}>\gamma_T)P_{los} +\mathbb{P}_{nlos}(\gamma_{nlos}>\gamma_T) (1-P_{los}) \\
&= \left(1-\frac{\tilde{\gamma}(m_{los},m_{los}\frac{\gamma_T}{\gamma_{los}})}{\Gamma(m_{los})}\right) P_{los}  \\
&\; \; \; \; \; \;\;\;\;\;\;\;\;\; + \left(1-\frac{\tilde{\gamma}(m_{nlos},m_{nlos}\frac{\gamma_T}{\gamma_{nlos}})}{\Gamma(m_{nlos})}\right) P_{nlos}. \\
\end{aligned}
\label{eq:coverage_probability}
\end{equation}
Here $\mathbb{P}_{los}$ is the complementary cdf of the Nakagami-$m$ fading power for LoS and $\mathbb{P}_{nlos}$ is the cdf of Nakagami-$m$ fading power for NLoS conditions. $\gamma_{los}$ and $\gamma_{nlos}$ are obtained by the SNR \eqref{eq:received_snr} with appropriate path loss exponent. The transmit power $P_t=30\;\si{\dBm}$, noise floor = $-100\; \si{\dBm}$ for a signal of bandwidth 20 MHz, $\gamma_T=-5\;\si{\dB}$   
\begin{figure}[!t]
\centering
\includegraphics[width=0.65\linewidth]
{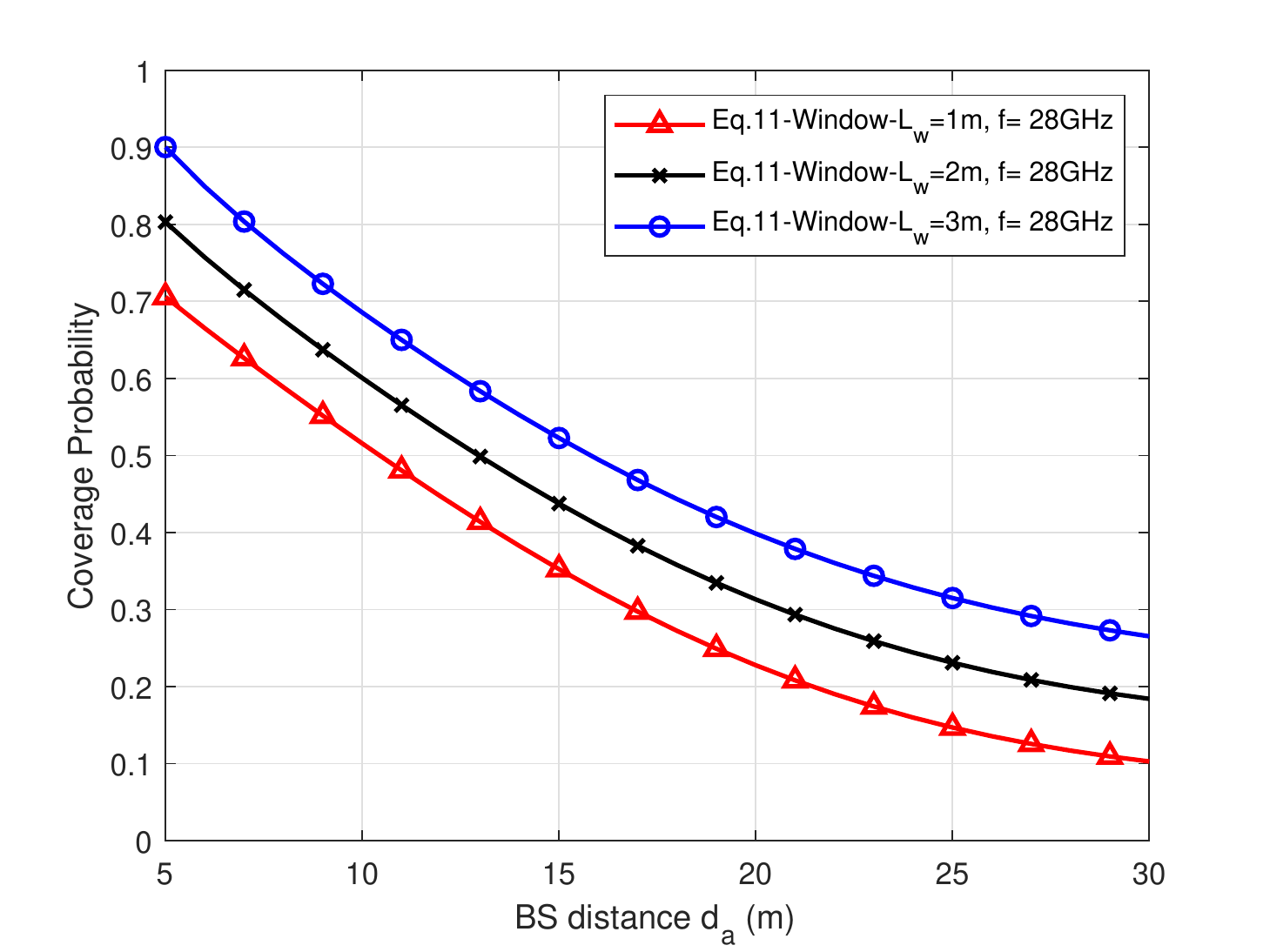}
         \caption{Coverage probability as a function of the BS distance for different window sizes. Frequency $f=28\si{\GHz}$, room dimension $L_r=20m$ and MS distance $d_n=20m$, threshold SNR $\gamma_t=$  5\;\si{\dB} for transmit power $P_t = 30\;\si{\dBm}$ , noise floor $= -100\;\si{\dBm}$. }
         \label{fig:coverage_probability}
\end{figure}

\section{Conclusion}
Recent propagation measurement campaigns have shown that the ratio of a window size to the overall building surface area, affects the link quality of a wireless link in an outdoor-to-indoor propagation scenario. In this paper, we used this fact to analyze the communication link quality in a UAV-assisted emergency network. We proposed a model that captures the relative position of the UAV-BS outside the building, the window size, the building size, and the transmit frequency of the link. This initial model focuses on a single room, but can easily be extended to multiple rooms. Based on the model and employing careful approximations, we derived a closed-form expression for the LoS probability of the link between the UAV-BS and MS inside the building. By analyzing the frequency dependence of the LoS probability, we observed that the 5G-mmWave bands offer significantly higher LoS probability compared to 5G-sub6 bands. Also, for 5G-mmWave bands, the LoS probability is directly proportional to the size of the window and inversely proportional to the smaller of the BS-to-building distance and the building size. The quality of the BS-to-MS wireless link is analyzed from the perspective of the coverage probability which is shown to be a function of the LoS probability. The dependence of the coverage probability on BS location, window size, and building size is also shown with short BS-to-building distance and large windows being favorable for better coverage. These results can be used to determine optimal UAV-BS positions and the number of UAVs needed to achieve desired reliability and performance.  
\section{Acknowledgments}
The authors express their gratitude to Prof. Harpreet S. Dhillon for his technical inputs. The statements, findings, conclusions, and recommendations are those of the author(s) and do not necessarily reflect the views of the National Institute of Standards and Technology (NIST) Public Safety Communications Research (PSCR) Division’s Public Safety Innovation Accelerator Program (PSIAP) or the U.S. Department of Commerce.

\bibliography{refs.bib}
\bibliographystyle{IEEEtran}

\end{document}